\documentclass[11pt,dvips]{article}
\usepackage{graphicx}
\usepackage{amsmath}
\usepackage{amssymb}
\usepackage{latexsym}
\usepackage{color}
\begin{document}
\thispagestyle{empty}
\begin{center}

\vspace{1.8cm}

%%%%%%%%%%%%%%%%%%%%%%%%%%%%%%%%%%%%%%%%%%%%%%%%%%%%%%%%%%%%%%%%%%%%%%%%%%%%%%%%%%%%%%%%%%%%%%%%%%%%%%%%%%%%%%%%%%%%%%%
 {\bf Quantum discord for multipartite coherent states interpolating between Werner and
Greenberger-Horne-Zeilinger states }\\
 %%%%%%%%%%%%%%%%%%%%%%%%%%%%%%%%%%%%%%%%%%%%%%%%%%%%%%%%%%%%%%%%%%%%%%%%%%%%%%%%%%%%%%%%%%%%%%%%%%%%%%%%%%%%%%%%%%%%%

\vspace{1.5cm}

{\bf M. Daoud}$^{a}${\footnote { email: {\sf
m$_{-}$daoud@hotmail.com}}} and {\bf R. Ahl Laamara}$^{b,c}$
{\footnote { email: {\sf ahllaamara@gmail.com}}}

\vspace{0.5cm}
$^{a}${\it Department of Physics , Faculty of Sciences, University Ibnou Zohr,\\
 Agadir ,
Morocco}\\[1em]

$^{b}${\it LPHE-Modeling and Simulation, Faculty  of Sciences,
University
Mohammed V,\\ Rabat, Morocco}\\[1em]

$^{c}${\it Centre of Physics and Mathematics,
CPM, CNESTEN,\\ Rabat, Morocco}\\[1em]

\vspace{3cm} {\bf Abstract}
\end{center}
\baselineskip=18pt
\medskip

 The quantum discord is used as measure
of quantum correlations for two families of multipartite coherent
states. The first family interpolates between generalized ${\rm
GHZ}$ states and generalized Werner states. The second one is an
interpolation between generalized ${\rm GHZ}$ and the ground state
of the multipartite quantum system. Two inequivalent ways to split
the system in a pair of qubits are introduced. The explicit
expressions of quantum quantum discord in multipartite coherent
states are derived. Its evaluation uses the Koashi-Winter relation
in optimizing the conditional entropy. The temporal evolution of
quantum correlations (quantum discord and entanglement) is also
discussed.

\newpage
\section{Introduction}

The characterization of quantum correlations  in a bipartite or
multipartite quantum system has been intensively investigated in the
context of quantum information science \cite{Horodecki-RMP-2009,
Guhne}. Different methods for quantifying these correlations were
reported in many works (for a recent review see \cite{vedral}).
Entanglement, one of the prominent of quantum correlations,
constitutes a typical resource to manage information in several ways
\cite{NC-QIQC-2000,Vedral-RMP-2002}. In this respect, a considerable
attention has been paid to develop a quantitative theory of
entanglement (concurrence, entanglement of formation and linear
entropy \cite{Rungta,Ben3,Wootters,Coffman}). Entanglement was for a
long time viewed as synonymous with quantum correlation. However,
some recent studies showed that entanglement is only a special kind
of quantum correlations. Indeed, unentangled quantum states can also
possess quantum correlations which play a relevant role in improving
quantum communication and information protocols better than their
classical counter-parts \cite{Knill}-\cite{Datta-PRL100-2008}.
Therefore, as the non-classicality of correlations present in
quantum states is not due solely to the presence of entanglement,
there was a need of a measure to characterize and quantify the
non-classicality or quantumness of correlations which goes beyond
entanglement.  The first attempt in this direction was made in the
works \cite{Vedral-et-al} and \cite{Ollivier-PRL88-2001} where the
authors concluded that when entanglement is subtracted from total
quantum correlation, there remain correlations that are not entirely
classical of origin. Nowadays, it is commonly accepted that the most
promising candidate to measure quantum correlations is the so-called
quantum discord.  It has attracted considerable attention and
continues to be intensively investigated in many contexts such as
quantum decoherence, quantum computation and phase transition  as
well as other related fields (a detailed list of references can be
found in \cite{vedral}).
%The explicit computation of quantum discord
%requires an optimization process over all possible quantum
%measurement. This makes the evaluation of quantum discord in general
%a very difficult task.
Many efforts were deployed to get a closed analytical expression of
quantum discord. Only some partial results were derived for some
special two-qubit systems
\cite{Vedral-et-al,Ollivier-PRL88-2001,Luo08,Mazhar10}. The
extension of  the notion of the discord to continuous variable
systems and the discussion of  its properties were considered in
\cite{Giorda}. It must be emphasized that there is now an intense
recent research activity to demonstrate experimentally the
advantages of the use of quantum discord without entanglement in
quantum protocols. In a recent work \cite{Dakic}, the authors
experimentally show, using a variety of polarization-correlated
photon pairs, non-zero quantum discord is the optimal resource for
remote state preparation. More interestingly, it is found that
unentangled state with non-zero quantum discord provides better
performance than entangled state by achieving a higher fidelity
(which is directly related to quantum discord) in remote state
preparation.\\

In this paper we shall be mainly concerned with the pairwise quantum
correlations, especially quantum discord, present in multipartite
coherent states. It must be noticed that the entangled coherent
states have received a special attention in the last two decades
(for a recent review, see \cite{Sanders3}). Therefore, paralleling
the treatment of entanglement in coherent states systems, the main
of this work is to investigate the pairwise quantum discord present
in multipartite nonorthogonal states. In fact, entangled coherent
states, which are typical examples of entangled nonorthogonal
states, have attracted much attention in the literature. One may
quote  for instance entangled Glauber coherent states
\cite{Sanders}, $SU(2)$ and $SU(1,1)$ entangled coherent states
\cite{Wang}. The term 'entangled coherent state' was introduced by
Sanders in a study concerning production of entangled coherent
states by using a nonlinear Mach-Zehnder interferometer
\cite{Sanders,Sanders2}. Entangled coherent states were initially
treated as bimodal states but later generalized to superpositions of
multimode coherent states \cite{Jex,Zheng,Wang1}. Generalizations to
multimode systems allow the intricacies of multipartite entanglement
to become manifest in entangled coherent states, as for instance in
{\rm GHZ} (Greenberger-Horne-Zeilinger), {\rm W} (Werner) states
\cite{Jeong1,Li} and entangled coherent state versions of cluster
states \cite{Munhoz,Wang-WF,Becerra}. In this sense, the
characterization
 of the quantum discord in nonorthogonal
multipartite states constitutes an important issue which deserves
the same degree of attention as entanglement.\\

This paper is organized as follows. In the first section, we
introduce a special instance of superpositions of multipartite
coherent states. They involve Weyl-Heisenberg, $SU(2)$ and $SU(1,1)$
coherent states labeled by a single complex parameter $z$. We shall
focus on balanced superpositions which are symmetric or
antisymmetric under the parity transformation $ z \longrightarrow -
z$. The symmetric superposition is interpolating between the ground
state of the multipartite system and ${\rm GHZ}$ state. The
antisymmetric superposition provides a continuous interpolation
between the Werner state and  ${\rm GHZ}$ state. To study the
bipartite quantum correlations present in the system, two different
qubit mapping are considered. Section 3  is devoted to the
derivation of the explicit expression of the quantum discord. The
optimization over all the measurement is performed by combining the
purification method and the Koashi-Winter relation . Finally, in the
last section we discuss the dynamical evolution of entanglement and
quantum discord under a dephasing channel. Concluding remarks close
this paper.

%%%%%%%%%%%%%%%%%%%%%%%%%%%%%%%%%%%%%%%%%%%%%%%%%%%%%%%%%%%%%%%%%%%%%%%%
\section{ Multipartite coherent states and qubit mapping}
%%%%%%%%%%%%%%%%%%%%%%%%%%%%%%%%%%%%%%%%%%%%%%%%%%%%%%%%%%%%%%%%%%%%%%%%
For several quantum systems, the coherent states can be obtained by
exploiting the structure relations of the relevant dynamical group
structure $G$. Therefore, any quantum state in the corresponding
Hilbert space can be expanded as a sum of the coherent states
associated with the group $G$. In this respect, a widespread
interest was devoted to the theoretical as well experimental studies
of superpositions of coherent states. Evidences of such
superpositions first appeared in a study of a certain type of
nonlinear Hamiltonian evolution in \cite{Milburn1,Milburn2}. Also, a
detailed analysis concerning  the manifestation of superpositions of
coherent states was reported in \cite{Yurke1,Yurke2} (see also
\cite{Buzek}). However, it is important to stress that
superpositions of coherent states are experimentally difficult to
produce, and fundamentally this could be due to extreme sensitivity
to environmental decoherence. Some experimental efforts to create
superpositions of coherent states were reported in \cite{Brune}.

%%%%%%%%%%%%%%%%%%%%%%%%%%%%%%%%%%%%%%%%%%%%%%%%%%%%%%%%%%%%%%%%%%%%%%%%%%%%%
\subsection{Basic of coherent states}
%%%%%%%%%%%%%%%%%%%%%%%%%%%%%%%%%%%%%%%%%%%%%%%%%%%%%%%%%%%%%%%%%%%%%%%%%%%%%

A system of coherent states is defined to be a set $\{ \vert z
\rangle; z \in {\cal D} \}$ of quantum states, in some  Hilbert
space ${\cal H}$, parameterized by the variable $z$ belonging to the
set ${\cal D}$ such that: $ z \longrightarrow \vert z \rangle$ is
 continuous and the system is over-complete; i.e.
\begin{equation}\label{mesure}
 \int \vert z \rangle \langle  z \vert~ d\mu(z)= I_{{\cal H}}.
\end{equation}
The continuity and the resolution of the unity (\ref{mesure}) form
the minimal set of requirements to define a system of coherent
states. They are not orthogonal to each other with respect to the
positive measure in (\ref{mesure}). It should be noticed that the
explicit form of the coherent state $\vert z \rangle$ in not
required for our purpose. However, to illustrate some interesting
limiting situations and for the sake of completeness, we consider
the familiar sets of Perelomov coherent states \cite{Perelomov}
associated with Weyl-Heisenberg, $SU(2)$ and $SU(1,1)$ groups (the
extension to other groups and other coherent states definitions  is
straightforward). A simple way to deal with these symmetries in a
unified scheme can be achieved by using  the so-called generalized
Weyl-Heisenbeg algebra ${\cal A}$ (see \cite{daoud} and references
therein) spanned by an annihilation operator $a^-$, a creation
operator $a^+$ and a number operator $N \not= a^+ a^-$ satisfying
the structure relations
\begin{eqnarray}
[a^- , a^+] =  G(N), \quad [N, a^{-}] = - a^{-}, \quad [N, a^{+}] =
+ a^{+}, \label{generalized W-H algebra}
\end{eqnarray}
with
\begin{eqnarray}
a^+ = (a^-)^{\dagger}, \quad N = N^{\dagger}, \quad G(N) = F(N+1) -
F(N), \label{hermiticity relations}
\end{eqnarray}
where the $F$ structure function characterizes the deviation from
the usual harmonic oscillator. A representation of  the algebra
${\cal A}$ which extends that of Weyl-Heisenberg, is defined through
 the actions
    \begin{eqnarray}
    && a^-\vert n \rangle = \sqrt{F(n)}    \vert n - 1 \rangle, \quad a^-\vert 0 \rangle = 0,
    \label{rep1}  \\
    && a^+\vert n \rangle = \sqrt{F(n+1)}  \vert n + 1 \rangle, \quad N \vert n \rangle = n \vert n \rangle
    \label{rep3}
    \end{eqnarray}
on the Hilbert space spanned by the eigenvectors of $N$. This
formally defines a representation of ${\cal A}$.  Note that $ a^+
a^- = F(N),$ a relation that generalizes $N = a^+ a^-$ for the
harmonic oscillator. The operator $F(N)$ can be considered as the
Hamiltonian for a quantum system. This provides a physical
significance to the structure function $F$. The representation
(Fock-Hilbert) space ${\cal H}$ is generated by the orthonormal set
$ \{ | n \rangle : n \ {\rm ranging} \}$. The dimension of the
representation afforded by (\ref{rep1}) and (\ref{rep3}) is
controlled by the positiveness of the eigenvalues $F(n)$ of the
operator $F(N)$. The usual Weyl-Heisenberg algebra is recovered for
$$ F(N) = N. $$
The algebra ${\cal A}$ coincides with $su(1,1)$ Lie algebra for
$$F(N) = N(2k-1+N)$$
where the number $k$, which acquires discrete values
$k=\frac{1}{2},1,\frac{3}{2},2,\ldots$, characterizes the $SU(1,1)$
discrete-series representations. The algebra ${\cal A}$  reproduces
the spin algebra $su(2)$ when
$$F(N) = N(2j+1-N),$$
and the creation and annihilation generators coincides with the
usual raising and lowering operators acting in the spin
representation of
dimension $2j+1$.\\

The Perelomov coherent states for the algebra ${\cal A}$ can be
constructed via a simple strategy based on the use of a
Fock-Bargmann space associated with ${\cal A}$ \cite{daoud}. This is
sketched briefly as follows. Let us look for states in the form
    \begin{eqnarray}
    \vert z  \rangle =  \sum_{n} a_n z^n \vert n \rangle, \quad a_n \in \mathbb{R}, \quad z \in \mathbb{C},
    \label{forme des ec}
    \end{eqnarray}
where the sum on $n$ is finite or infinite according to as ${\cal
A}$ admits a finite- or infinite-dimensional representation. The
$a_n$ coefficients can then be determined from the correspondence
rules
    \begin{eqnarray}
    \vert n \rangle \longrightarrow a_{n} z^{n}, \qquad
    a^-\longrightarrow \frac{d}{dz}
    \label{correspondance pour Perelomov}
    \end{eqnarray}
applied to relations (\ref{rep1}) and (\ref{rep3}).  This analytical
realization  leads to the following recurrence relation
    \begin{eqnarray}
    n a_{n}= \sqrt{F(n)}a_{n-1},
    \label{recur}
    \end{eqnarray}
which can be iterated to give
    \begin{eqnarray}
    a_{n}= \frac{\sqrt{F(n)!}}{n!},
    \label{recur2}
    \end{eqnarray}
(by taking $a_0 = 1$). In Eq.~(\ref{recur2}), the generalized
factorials are defined by
$$
F(0)! = 1, \qquad F(n)! = F(1) F(2) \ldots F(n).
$$
This yields the following normalized coherent states
\begin{equation}
    \vert z \rangle = N(\vert z \vert)\sum_{n}
  \frac{\sqrt{F(n)!}}{n!} z^n  \vert n \rangle
   \label{cs-main}
\end{equation}
where the normalization factor is
$$\bigg[N(\vert z \vert)\bigg]^{-2} = \sum_{n}   \frac{F(n)!}{(n!)^2} \vert z\vert^{2n}.$$
subject to convergence. They satisfy $\vert z \rangle = \exp( z a^+)
\vert 0 \rangle$ and are thus coherent states in the Perelomov
sense. For the usual harmonic oscillator ($F(n) = n$), the equation
(\ref{cs-main}) reads
\begin{equation}
\vert z \rangle =e^{-\frac{\vert z \vert^2}{2}} \sum_{n=0}^{\infty}
\frac{z^n}{\sqrt{n!}} \vert n \rangle. \label{glauber-cs}
\end{equation}
The overlap between two Glauber coherent states is
\begin{equation}
\langle z_1 \vert z_2 \rangle = \exp(-\frac{1}{2} (\vert z_1 \vert^2
+ \vert z_2 \vert^2  - 2 \bar z_1 z_2)).\label{overlap-glauber}
\end{equation}
 The standard set of the $SU(2)$ coherent state is
obtained from (\ref{cs-main}) for $F(n) = n(2j+1-n)$. They  are
given by
\begin{equation}
| z\rangle = (1+ \vert z \vert^{2})^{-j} \sum_{n=0}^{2j} \left[
\frac{(2j)!}{n!(2j-n)!} \right]^{1/2} z^{n} |n\rangle.
\label{su2-cs}
\end{equation}
The parameter $z$ can acquire any complex value. The $SU(2)$
coherent states are normalized but they are not orthogonal to each
other:
\begin{equation}
\langle z_{1}|z_{2}\rangle = (1+|z_{1}|^{2})^{-j}
(1+|z_{2}|^{2})^{-j} (1 + \bar z_{1} z_{2})^{2j}.
\label{overlap-su2}
\end{equation}
Finally, for $F(n)= n(2k-1+N)$, one gets the $SU(1,1)$ coherent
states
\begin{equation}
|z \rangle = (1-|z|^{2})^{k} \sum_{n=0}^{\infty}
\left[\frac{(2k-1+n)!}{n!(2k-1)!}\right]^{1/2} z^{n} |n\rangle.
\label{su11-cs}
 \end{equation}
where the complex variable $z$ belongs to the unit disc $\{ |z| \in
\mathbb{C}, \quad |z|<1\}$.
 The kernel of two $SU(1,1)$ coherent states reads
\begin{equation}
\langle z_{1}|z_{2}\rangle = (1-|z_{1}|^{2})^{k} (1-|z_{2}|^{2})^{k}
(1 - \bar z_{1} z_{2})^{-2k}. \label{ovelap-su11}
\end{equation}

%%%%%%%%%%%%%%%%%%%%%%%%%%%%%%%%%%%%%%%%%%%%%%%%%%%%%%%%%%%%%%%%%%%%%%%%%%%
\subsection{Superpositions of multipartite coherent states}
%%%%%%%%%%%%%%%%%%%%%%%%%%%%%%%%%%%%%%%%%%%%%%%%%%%%%%%%%%%%%%%%%%%%%%%%
The over-completeness relation makes possible the expansion of an
arbitrary state of the Hilbert space ${\cal H}$ in terms of the
coherent states of the system under consideration. It follows that
when one considers a collection of $n$ noninteracting identical
particles, the whole Hilbert space is a tensor product and any
multipartite coherent state can be written as a superposition of
tensorial products of the form $\vert z_1 \rangle\otimes\vert z_2
\rangle\cdots \otimes\vert z_n \rangle \equiv \vert z_1, z_2, \cdots
z_n \rangle$. In fact, the resolution of identity allows us to
expand any state $\vert \psi \rangle $ in the space ${\cal H}
\otimes{\cal H} \otimes \cdots {\cal H}$ as
\begin{equation}\label{ncs}
 \vert \psi \rangle  = \int ~ d\mu(z_1) ~ d\mu(z_2)\cdots ~ d\mu(z_n)\vert z_1, z_2, \cdots z_n  \rangle \langle  z_1, z_2, \cdots z_n
 \vert \psi \rangle
\end{equation}
reflecting that any multipartite state can be viewed as a
superposition of the coherent states $\vert z_1, z_2, \cdots z_n
\rangle$. The multipartite  state (\ref{ncs}) can be reduced to a
sum if the function $ \psi (z_1, z_2, \cdots z_n) = \langle  z_1,
z_2, \cdots z_n
 \vert \psi \rangle$  can be expressed as a sum of delta functions.
 Indeed, setting
$$ \psi (z_1, z_2, \cdots z_n) = (\delta(z - z_1) + e^{im \pi} \delta(z + z_1)) \delta(z_1 - z_2) \delta(z_2 - z_3)\cdots \delta(z_{n-1} - z_n),$$
one gets the following balanced or equally weighted  superposition
of multipartite coherent states
\begin{equation}
\vert \psi \rangle \equiv \vert z, m, n \rangle ={\cal N}(\vert z
\rangle\otimes \vert z \rangle\otimes \cdots\otimes \vert z \rangle
+e^{im\pi}|- z \rangle\otimes |- z \rangle\otimes \cdots\otimes |- z
\rangle) \label{eq:main}
\end{equation}
where $m \in \mathbb{Z}$. The normalization factor ${\cal N}$ is
$$ {\cal N} = \big[ 2 + 2 p^n \cos m \pi\big]^{-1/2}$$
where  $p$ denotes the overlap between the states $\vert z \rangle$
and $\vert - z \rangle$. It is given
$$ p = \langle z \vert - z \rangle.$$
It is real as it can be verified from (\ref{cs-main}). For
Weyl-Heisenberg, $SU(2)$ and $SU(1,1)$ coherent states, the quantity
$p$ is obtainable from the expressions (\ref{overlap-glauber}),
(\ref{overlap-su2}) and (\ref{ovelap-su11}), respectively. Two
interesting limits of superpositions of the form (\ref{eq:main})
arise when $ p \rightarrow 0$ and $ p \rightarrow 1$.

We first consider the asymptotic limit $p \rightarrow 0 $. In this
limit  the two states $|z \rangle $ and $|- z \rangle $ approach
orthogonality, and an orthogonal basis can be constructed such that
$\vert {\bf 0}\rangle\equiv \vert z \rangle$ and $\vert{\bf
1}\rangle \equiv \vert - z \rangle$. Thus, the state $ \vert z , m
,n \rangle$ approaches a multipartite state of ${\rm GHZ}$ type
\begin{equation}
\vert z , m, n \rangle \sim \vert {\rm GHZ}\rangle_n = \frac
1{\sqrt{2}}(\vert {\bf 0}\rangle \otimes |{\bf 0}\rangle \otimes
        \cdots \otimes\vert {\bf 0}\rangle
    +e^{i m \pi}\vert {\bf 1}\rangle \otimes
    \vert {\bf 1}\rangle \otimes \cdots \otimes
\vert {\bf 1}\rangle).\label{GHZ}
\end{equation}
In the situation where $p \rightarrow 1$ (or $ z \rightarrow 0$ ),
one should distinguish separately the cases $m = 0 ~({\rm mod}~2)$
and $m = 1 ~({\rm mod}~2)$. For $m$ even, the multipartite
superposition (\ref{eq:main}) reduces to ground state
\begin{equation}
\vert 0 , 0 ~({\rm mod}~ 2) , n \rangle \sim  \vert 0\rangle
\otimes\vert 0 \rangle \otimes \cdots \otimes \vert 0 \rangle,
\end{equation}
and  for $m$ odd, the state $\vert z , 1 ~({\rm mod}~ 2) ,n \rangle$
reduces to  a multipartite state of Werner type~\cite{Dur00}
\begin{equation}
\vert 0 ,  1 ~({\rm mod}~ 2), n \rangle \sim \vert\text{\rm
W}\rangle_n
    = \frac{1}{\sqrt{n}}(\vert 1\rangle \otimes\vert 0 \rangle \otimes \cdots\otimes
       \vert0\rangle  +\vert 0\rangle \otimes\vert 1\rangle \otimes\ldots\otimes \vert0\rangle
      +\cdots
   + \vert 0\rangle \otimes\vert 0\rangle  \otimes \cdots\otimes \vert 1\rangle)~.
\label{Wstate}
\end{equation}
Here $\vert n\rangle $ $(n=0,1)$ denote the Fock-Hilbert states.

It follows that the states $\vert z, m = 0 ~({\rm mod}~2), n\rangle$
interpolate between states of ${\rm GHZ}$ type $(p \rightarrow 0)$
and the separable state $\vert 0\rangle \otimes\vert 0 \rangle
\otimes \cdots \otimes \vert 0 \rangle$ $(p \rightarrow 1)$. In
other hand, the states $\vert z , m = 1 ~({\rm mod}~2), n\rangle$
may be viewed as interpolating between states of ${\rm GHZ}$ type
$(p \rightarrow 0)$ and states of Werner type $(p \rightarrow 1)$.

To close this subsection, it is important to emphasize that the main
concern in investigating the properties of superpositions of
 coherent states is how to produce such states. Their experimental production is
 fundamentally difficult to achieve. This is especially due to
 extreme sensitivity to environmental decoherence.
Experimental efforts to create superpositions of
 coherent states with the present day technology, are encouraging.
 Indeed,
superpositions of weak coherent states with opposite phase,
resembling to a small "Schr\"odinger's cat" state (or
"Schr\"odinger's kitten"), were produced by photon subtraction from
squeezed vacuum \cite{Ourjoumtsev}. Also, the experimental
generation of arbitrarily large squeezed Schrodinger cat states,
using homodyne detection and photon number states (two photons) as
resources was reported in \cite{Ourjoumtsev}.  Very recently,
creation of coherent state superpositions, by subtracting up to
three photons from a pulse of squeezed vacuum light, is reported in
\cite{Gerrits}. The
 mean photon number of  such coherent states produced by
three-photon subtraction is of 2:75. The production of cat states
especially ones of high amplitude or mean number of photons remains
an experimental challenge. Considering the fast technical progress
and the increasing number of groups working in this field, we expect
that the generation of cat states (and Bell states) is a goal that
is achievable in the near future.

%%%%%%%%%%%%%%%%%%%%%%%%%%%%%%%%%%%%%%%%%%%%%%%%%%%%%%%%%%%%%%%%
\subsection{Qubit mapping}
%%%%%%%%%%%%%%%%%%%%%%%%%%%%%%%%%%%%%%%%%%%%%%%%%%%%%%%%%%%%%%%%%%%
%%%%%%%%%%%%%%%%%%%%%%%%%%%%%%%%%%
\subsubsection{Pure  states}
%%%%%%%%%%%%%%%%%%%%%%%%%%%%%%%%%%
To study the bipartite quantum correlations present in
(\ref{eq:main}), the whole system can be partitioned in two
different ways. We first consider bipartite splitting of the
multipartite system, i.e., splitting the entire system into two
subsystems, one subsystem containing any~$k$ $(1\le k\le n-1)$
particles and the other containing the remaining $n-k$ particles.
Accordingly, one writes the state $\vert z, m, n \rangle$ as
\begin{equation}\label{partition1}
 \vert z, m, n
\rangle = {\cal N} (\vert z  \rangle_k \otimes \vert z \rangle_{n-k}
+ e^{im\pi } \vert -z  \rangle_k \otimes \vert -z \rangle_{n-k})
\end{equation}
where
$$ \vert \pm z  \rangle_l =  \vert \pm z \rangle_1\otimes
\vert \pm z \rangle_2\otimes \cdots\otimes \vert \pm z \rangle_l ,
\qquad l = k , n-k.
$$
The multi-particle state $\vert z, m , n \rangle$ can be expressed
as a state of two logical qubits. For this end, we introduce, for
the first subsystem, the orthogonal basis $\{ \vert 0 \rangle_k ,
\vert 1 \rangle_k\}$ defined as
\begin{equation}\label{base1}
\vert 0 \rangle_k = \frac{ \vert z  \rangle_k +  \vert -z
\rangle_k}{\sqrt{2(1 + p^k)}}
   \qquad \vert 1 \rangle_k = \frac{\vert z \rangle_k -  \vert - z
\rangle_k}{{\sqrt{2(1- p^{k})}}}.
\end{equation}
 Similarly, we introduce, for the second subsystem, the orthogonal
basis $\{ \vert  0 \rangle_{n-k} , \vert  1 \rangle_{n-k}\}$ given
by
\begin{equation}\label{base2}
\vert 0 \rangle_{n-k} = \frac{ \vert z  \rangle_{n-k} +  \vert -z
\rangle_{n-k}}{\sqrt{2(1 + p^{n-k})}}
   \qquad \vert 1 \rangle_{n-k} = \frac{\vert z \rangle_{n-k} -  \vert - z
\rangle_{n-k}}{{\sqrt{2(1- p^{n-k})}}}.
\end{equation}
Reporting the equations (\ref{base1}) and (\ref{base2}) in
(\ref{partition1}), one has the explicit form of the pure state
$\vert z, m, n \rangle$ in the basis $\{ \vert 0 \rangle_{k} \otimes
\vert 0 \rangle_{n-k} ,
 \vert 0 \rangle_{k} \otimes \vert 1 \rangle_{n-k} , \vert  1 \rangle_{k}
 \otimes \vert 0 \rangle_{n-k} , \vert  1 \rangle_{k} \otimes \vert  1
 \rangle_{n-k}\}$. It is given by
\begin{equation}
\vert z, m, n \rangle = \sum_{\alpha= 0,1} \sum_{\beta= 0,1}
C_{\alpha,\beta} \vert \alpha \rangle_k \otimes \vert \beta
\rangle_{n-k}\label{mapping1}
\end{equation}
where the coefficients $C_{\alpha,\beta}$ are
$$ C_{0,0} = {\cal N}(1 + e^{im\pi}) a_{k}a_{n-k}  , \qquad  C_{0,1} =  {\cal N} (1 -e^{im\pi}) a_{k}b_{n-k} $$
$$ C_{1,0} = {\cal N} (1 - e^{im\pi}) a_{n-k}b_{k}  , \qquad  C_{1,1} =  {\cal N} (1 + e^{im\pi}) b_{k}b_{n-k}. $$
in term of the quantities
$$ a_l =\sqrt{\frac{1+p^l}{2}} , \qquad b_l = \sqrt{\frac{1-p^l}{2}} \qquad {\rm for} ~ l = k, n-k$$
involving the overlap $p$ between two coherent states of  equal
amplitude and opposite phase.

\subsubsection{Mixed states}

The second partition can be realized by considering  the bipartite
reduced density matrix $\rho_{kl}$ which is obtained by tracing out
all other subsystems except ones labeled by the indices $k$ and $l$.
There are $n(n-1)/2$ different density matrices $\rho_{kl}$. It is
simple to see that all the reduced density matrices $\rho_{kl}$ are
identical. Therefore, it is sufficient to consider $\rho_{12}$ and
to generalize from this case. Then, by tracing out systems
$3,4,\ldots,n$ in the state $\vert z , m, n \rangle$, we obtain the
reduced density matrix describing the particles or modes 1 and 2 as
\begin{eqnarray}
\rho_{12} &=&\text{Tr}_{3,4,\ldots,n}(\vert z, m, n\rangle \langle
z, m,n |)  \nonumber \\
&=&{\cal N}^2(\vert z , z \rangle \langle z , z \vert +\vert - z , -
z \rangle \langle - z , - z | + e^{i m \pi } q |- z , - z \rangle
\langle z , z  \vert +e^{-i m \pi }q\vert z , z \rangle \langle - z,
- z \vert )  \label{rho12}
\end{eqnarray}
with $q \equiv p^{n-2}$. To study the correlations of the system
described by the density matrix $\rho_{12}$, we convert it into a
two-qubit system. Thus, we choose  an orthogonal pair $\{\vert {\bf
0}\rangle ,\vert {\bf 1}\rangle \}$ as
\begin{equation}
\vert z \rangle \equiv  a \vert {\bf 0} \rangle + b  \vert {\bf 1}
\rangle ~,\; \vert - z \rangle\equiv a \vert {\bf 0} \rangle - b
 \vert {\bf 1} \rangle,\label{base}
\end{equation}
where $$a = \sqrt{\frac{1+p}{2}} \qquad b = \sqrt{\frac{1-p}{2}}.$$
The logical   qubits $\vert {\bf 0} \rangle $  and $\vert {\bf 1}
\rangle $  given by
\begin{equation}
\vert {\bf 0} \rangle =      \frac{1}{\sqrt{2+2p}}(\vert z \rangle +
\vert -z \rangle) ~  \qquad \vert {\bf 1} \rangle =
\frac{1}{\sqrt{2-2p}}(\vert z \rangle - \vert -z
\rangle).\label{qubit-mixed}
\end{equation}
coincide with even and odd coherent states, respectively.
Substituting the equation (\ref{base}) into (\ref{rho12}), we obtain
the density matrix
%({\bf attention: utiliser a,b ou $a_p, b_p$})
\begin{equation}
\rho_{12} = {\cal N}^2 \left( \begin{smallmatrix} 2a^4(1+q\cos
m\pi)& 0
    & 0& 2a^2b^2(1+q\cos m\pi
)\\
0  & 2a^2b^2(1-q\cos m\pi ) & 2a^2b^2(1-q\cos m\pi
) & 0 \\
0  & 2a^2b^2(1-q\cos m\pi ) & 2a^2b^2(1-q\cos m\pi
) & 0 \\
2a^2b^2(1+q\cos m\pi ) & 0 & 0 & 2b^4(1+q\cos m\pi)
\end{smallmatrix}
\right) \label{rho12-matrix}
\end{equation}
in the basis $\{\vert{\bf 00}\rangle ,\vert{\bf 01}\rangle
,\vert{\bf 10}\rangle ,
    \vert{\bf 11}\rangle \}$. It is remarkable that the obtained
    density belongs to the set of the so-called $X$ states. It can be also written,
    in the Bloch representation,   as
\begin{equation}
 \rho_{12} = \sum_{\alpha = 0}^{3}\sum_{\beta = 0}^{3}
R_{\alpha,\beta} \sigma^{\alpha} \otimes \sigma^{\beta}
\end{equation}
where the correlation matrix $R$ is given by
\begin{equation}
R = \left(%
\begin{array}{cccc}
  1 & 0 & 0 & 2{\cal N}^2p(1+ q\cos m\pi) \\
  0 & 2{\cal N}^2(1-p^2) &  0 & 0 \\
  0 & 0 & -2{\cal N}^2(1-p^2)q\cos m\pi  & 0 \\
  2{\cal N}^2p( 1+ q \cos m\pi) & 0 & 0 & 2{\cal N}^2(p^2+ q\cos m\pi) \\
\end{array}%
\right)\label{R-matrix}
\end{equation}
and $\sigma^0$ and $\sigma^i$ ($i = 1,2,3$) stand for identity and
the usual Pauli matrices, respectively.

%%%%%%%%%%%%%%%%%%%%%%%%%%%%%%%%%%%%%%%%%%%%%%%%%%%%%%%%%%%%%%%%%%%%%
\section{Quantifying the quantum discord}
%%%%%%%%%%%%%%%%%%%%%%%%%%%%%%%%%%%%%%%%%%%%%%%%%%%%%%%%%%%%%%%%%%%%%%

So far, quantum discord has been calculated explicitly only for a
rather limited set of two-qubit quantum states and analytical
expressions for more general quantum states are not known. In this
section, we give another instance of states for which quantum
discord can be evaluated explicitly. Indeed, we shall derive the
explicit form of this kind of bipartite correlation in the state
(\ref{eq:main}) by making use of  the qubit mapping corresponding to
the partitioning schemes (\ref{partition1}) and (\ref{rho12})
discussed in the previous section.

%%%%%%%%%%%%%%%%%%%%%%%%%%%%%%%%%%%%%%%%%%%%%%%%%%%
\subsection{Definitions}
%%%%%%%%%%%%%%%%%%%%%%%%%%%%%%%%%%%%%%%%%%%%%%%%%%%

The quantum discord is defined as the difference between total
correlation and classical correlation. The total correlation is
usually quantified by the mutual information $I$
\begin{equation}\label{def: mutual information}
    I(\rho_{AB})=S(\rho_A)+S(\rho_B)-S(\rho_{AB}),
\end{equation}
where $\rho_{AB}$ is the state of a bipartite quantum system
composed of the subsystems $A$ and $B$, the operator
$\rho_{A(B)}={\rm Tr}_{B(A)}(\rho_{AB})$ is the reduced state of
$A$($B$) and $S(\rho)$ is the von Neumann entropy of a quantum state
$\rho$. Suppose that a positive operator valued measure (POVM)
measurement is performed on subsystem $A$. The set of POVM elements
is denoted by $\mathcal{M}=\{M_k\}$ with $M_k\geqslant 0$ and
$\sum_k M_k= \mathbb{I} $. In this paper, we deal with two-qubit
rank two states (with only two nonzero eigenvalues). In this case,
the generalized positive operator valued measurement is not
required. Indeed, it has been clearly shown in \cite{Hamieh,Galve}
(see also the recent review \cite{Modi}) that for a bipartite mixed
state of two qubits  of rank two , the optimal measurement giving
the quantum discord is a two element POVM. The elements of such POVM
are orthogonal projectors. Thus, it is natural to consider only
projective measurements for the subsystem $A$. The von Neumann
measurement (from now on just a measurement) on the subsystem $A$
project the system into a statistical ensemble $\{ p_{B,k} ,
\rho_{B,k}\}$ such that
$$\rho_{AB} \longrightarrow \frac{(M_k \otimes \mathbb{I})\rho_{AB}(M_k \otimes \mathbb{I})}{p_{B,k}}$$
where the measurement operation is written as \cite{Luo08}
\begin{eqnarray}
M_k = U \, \Pi_k \, U^\dagger \label{Eq:VNmsur}
\end{eqnarray}
with $\Pi_k = |k\rangle\langle k| ~ (k = 0,1)$ is the projector for
subsystem $A$  along the computational base $|k\rangle$,  $U \in
SU(2)$ is a unitary operator and
$$ p_{B,k} = {\rm tr}  \bigg[ (M_k \otimes \mathbb{I})\rho_{AB}(M_k \otimes \mathbb{I}) \bigg]. $$
The amount of information acquired about particle $B$ is then given
by
$$S(\rho_B)-\sum_k ~p_{B,k} ~S(\rho_{B,k}),$$
which depends on measurement $\mathcal{M}$. This dependence can be
removed by doing maximization over all the measurements, which gives
rise to the definition of classical correlation
\begin{eqnarray}
    C(\rho_{AB})& =\max_{\mathcal{M}}
    \Big[S(\rho_B)-\sum_k ~p_{B,k} ~S(\rho_{B,k})\Big] \nonumber \\
    & =S(\rho^B) - \widetilde{S}_{\rm min}
      \label{def: classical correlation}
\end{eqnarray}
where $\widetilde{S}_{\rm min}$  denotes the minimal value of the
conditional  entropy

\begin{equation}
\widetilde{S} =  \sum_k ~p_{B,k}
~S(\rho_{B,k}).\label{condit-entropy}
\end{equation}
It follows that quantum discord is then given by
\begin{equation} \label{def: discord}
    D(\rho_{AB})= I(\rho_{AB}) - C(\rho_{AB})
    =S(\rho_A)+\widetilde{S}_{\rm min}-S(\rho_{AB}).
\end{equation}
The main step in evaluating the quantum discord is the minimization
of conditional entropy  to get an explicit expression of the quantum
discord in the multipartite system (\ref{eq:main}).

%%%%%%%%%%%%%%%%%%%%%%%%%%%%%%%%%%%%%%%%%%%%%%%%%%%%%%%%%%%%%%%%%%%%%%%%%%%%%%%
\subsection{ Quantum discord for pure bipartite coherent states}
%%%%%%%%%%%%%%%%%%%%%%%%%%%%%%%%%%%%%%%%%%%%%%%%%%%%%%%%%%%%%%%%%%%%%%%%%%%%%%%%

According to the first partitioning scheme (\ref{partition1}), the
entire $n$ particles system is split into two components $A$,
containing $k$ particles, and $B$, containing $n-k$ particles. The
bipartite density state $\rho_{k,n-k} = \vert z, m, n \rangle
\langle  z, m, n \vert$ is pure and the conditional density is also
a pure state. This implies that the quantum conditional entropy is
zero. Then, the quantum discord for the pure state $\rho_{AB}\equiv
\rho_{k,n-k}$ coincides with the von Neumann entropy of the
subsystem $A$:
\begin{equation}
 D(\rho_{k,n-k}) = S(\rho_{k})
\end{equation}
where $\rho_{k}$ is the reduced density of the subsystem $A$. In
this scheme  the quantum discord can be computed easily. It is given
by
\begin{equation}
D(\rho_{k,n-k}) = - \lambda_+ \log_2 \lambda_+ - \lambda_- \log_2
\lambda_-
\end{equation}
in term of the eigenvalues of the reduced density matrix $\rho_{k}$
given by
\begin{equation}
\lambda_{\pm}= \frac{1}{2}\bigg( 1 \pm \sqrt{1 - {\cal C}^2_{k,n-k}}
\bigg)
\end{equation}
where ${\cal C}_{k,n-k}$ is the concurrence between the  subsystems
$A$ and $B$:
\begin{equation}
{\cal C}_{k,n-k} = 2 \vert C_{0,0} C_{1,1} - C_{1,0}C_{0,1} \vert
=\frac{\sqrt{1-p^{2k}}\sqrt{1-p^{2(n-k)}}}{1+p^n\cos
m\pi}\label{concurence1}
\end{equation}
where we used the mapping (\ref{mapping1}). Note that
 the entanglement of formation given by
\begin{equation}
E(\rho_{k,n-k}) = H \bigg(\frac{1}{2} + \frac{1}{2}\frac{p^k +
p^{n-k}\cos m\pi}{1 + p^n\cos m\pi}\bigg),\label{qdpure}
\end{equation}
is nothing but the von Neumann entropy of the subsystem $A$. Here
$H$ stands for the binary entropy $ H(x)=h(x)+h(1-x)$ with $ h\left(
x\right)  =-x\log_{2}x$. Hence, we have the following closed
relation between quantum discord and entanglement of formation
\begin{equation}
D(\rho_{k,n-k}) = E(\rho_{k,n-k}).
\end{equation}
This agrees with the fact that quantum discord and entanglement of
formation are identical for pure states and amount to the same set
of
correlations.\\
In the limiting case $p \rightarrow 0$, the state (\ref{partition1})
is of ${\rm GHZ}$ type having a maximal bipartite entanglement
$({\cal C}_{k,n-k} = 1)$ and we obtain
$$  D(\rho_{k,n-k})= 1.$$
The situation becomes different when $p \rightarrow 1$. In this
case, we have
$$ D(\rho_{k,n-k})= 0$$
for symmetric pure states (i.e. $m$ even) as expected (in this limit
the state (\ref{eq:main}) is a $n$ tensorial product of the ground
state $\vert 0 \rangle$). In the limit $p \rightarrow 1$, the
antisymmetric states (i.e. $m$ odd) are of Werner type. The
bipartite concurrence is
$${\cal C}_{k,n-k} =
\frac{2}{n}\sqrt{k(n-k)},$$ and the corresponding pairwise quantum
discord takes the form
$$ D(\rho_{k,n-k})=
H\bigg(\frac{1}{2} + \frac{1}{2}\frac{\vert n-2k\vert}{n} \bigg).$$
The quantum correlations (entanglement and quantum discord) in
multipartite Werner states vanish as $n$ becomes large.

%%%%%%%%%%%%%%%%%%%%%%%%%%%%%%%%%%%%%%%%%%%%%%%%%%%%%%%%%%%%%%%%%%%%%%%%%%%%%
\subsection{ Quantum discord for mixed bipartite coherent states}
%%%%%%%%%%%%%%%%%%%%%%%%%%%%%%%%%%%%%%%%%%%%%%%%%%%%%%%%%%%%%%%%%%%%%%%%%

%%%%%%%%%%%%%%%%%%%%%%%%%%%%%%%%%%%%%%%%%%%%%%%%%%%%%%%%%%%%%%%%%
\subsubsection{ Mutual information entropy}
%%%%%%%%%%%%%%%%%%%%%%%%%%%%%%%%%%%%%%%%%%%%%%%%%%%%%%%%%%%%%%%%%

The subsystems $A$ and $B$ of the subsection 3.1 correspond here to
the modes or particles 1 and 2 described by the density operator
$\rho_{12}$ (\ref{rho12-matrix}). The non vanishing eigenvalues of
the density matrix $\rho_{12}$ are
\begin{equation}
\lambda_{\pm} = \frac{1}{2} \frac{(1\pm p^2)(1 \pm q\cos(m\pi))}{1 +
p^n\cos(m\pi)},\label{lambda+-}
\end{equation}
and the joint entropy is
\begin{equation}\label{entropy12}
S(\rho_{12}) = h(\lambda_+) + h(\lambda_-) = H(\lambda_+).
\end{equation} The eigenvalues of the marginal $\rho_1 = {\rm Tr}_2
\rho_{12}$ are
$$\lambda_{1,\pm} = \frac{1}{2} \frac{(1\pm p)(1 \pm pq\cos(m\pi))}{1 + p^n\cos(m\pi)},$$
and the marginal entropy reads
\begin{equation}\label{entropy1}
S(\rho_{1}) = h(\lambda_{1,+}) + h(\lambda_{1,-}) =
H(\lambda_{1,+}).
\end{equation}
The eigenvalues of the marginal $\rho_2 = {\rm Tr}_1 \rho_{12}$ are
$$\lambda_{2,\pm} = \frac{1}{2} \frac{(1\pm p)(1 \pm pq\cos(m\pi))}{1 + p^n\cos(m\pi)},$$
and the corresponding entropy is given by
\begin{equation}\label{entropy2}
S(\rho_{2}) = h(\lambda_{2,+}) + h(\lambda_{2,-}) =
H(\lambda_{2,+}).
\end{equation}
Note that  for the mixed density under consideration the marginal
densities $\rho_1$ and $\rho_2$ are identical. The explicit form of
the mutual information (\ref{def: mutual information}) writes
\begin{equation}
    I(\rho_{12})= 2 H\bigg( \frac{1}{2} \frac{(1 + p)(1 + p^{n-1}\cos(m\pi))}{1 + p^n\cos(m\pi)}\bigg)
    -  H\bigg( \frac{1}{2} \frac{(1 + p^2)(1 + p^{n-2}\cos(m\pi))}{1 +
p^n\cos(m\pi)}\bigg).
\end{equation}

%%%%%%%%%%%%%%%%%%%%%%%%%%%%%%%%%%%%%%%%%%%%%%%%%%%%%%%%%%%%%%%%
\subsubsection{Conditional entropy}
%%%%%%%%%%%%%%%%%%%%%%%%%%%%%%%%%%%%%%%%%%%%%%%%%%%%%%%%%%%%%%%%
After computing the quantum mutual information, we  next compute the
classical correlation $C(\rho_{12}) \equiv C(\rho_{AB})$. This
requires an explicit expression of the unitary operator $U$
occurring in (\ref{Eq:VNmsur}). So, we parameterizes $U$ as follows
$$ U = \exp(\eta \sigma_+ - \bar\eta \sigma_-) \exp(i\phi \sigma_3) \qquad \eta \in \mathbb{C}, ~ \phi \in \mathbb{R}.$$
Using this parametrization, one can verify that the quantities
defined by
$$\langle \sigma_i \rangle_ k = \langle k \vert U^\dagger \sigma_i  U \vert k \rangle, \qquad i = 1,2, 3 \quad {\rm and} \quad k = 0,1$$
are given by
\begin{eqnarray}
\langle \sigma_3 \rangle_ k = (-)^k \frac{1-\bar \alpha \alpha
}{1+\bar \alpha \alpha},\qquad  \langle \sigma_1 \rangle_ k = (-)^k
\frac{\bar \alpha + \alpha}{1+\bar \alpha \alpha} \qquad , \langle
\sigma_2 \rangle_ k = i (-)^k \frac{\bar \alpha - \alpha}{1+\bar
\alpha \alpha}\label{val-moy-z}
\end{eqnarray}
where $ \alpha = - i \frac{\eta}{\sqrt{\bar\eta \eta}} \tan
\sqrt{\bar\eta \eta} $. They can be also written as
\begin{eqnarray}
\langle \sigma_3 \rangle_ k = (-)^k \cos \theta,\qquad  \langle
\sigma_1 \rangle_ k = (-)^k  \sin \theta \cos \varphi  \qquad ,
\langle \sigma_2 \rangle_ k = (-)^k \sin \theta \sin \varphi
\label{val-moy-theta}
\end{eqnarray}
where $\frac{\theta}{2}e^{i\varphi} = -i \eta$. The conditional
density operators $\rho_{B,k}  \equiv \rho_{2,k}$ are
\begin{eqnarray}
\rho_{2,k} = \frac{1}{p_{2,k}}\left(%
\begin{array}{cc}
  (1+ R_{03}) + (R_{30}+ R_{33})\langle \sigma_3\rangle_k & R_{11}\langle \sigma_1\rangle_k - i R_{22}\langle \sigma_2\rangle_k\\
 R_{11}\langle \sigma_1\rangle_k + i R_{22}\langle \sigma_2\rangle_k & (1- R_{03}) + (R_{30}- R_{33})\langle \sigma_3\rangle_k \\
\end{array}%
\right) \label{rho2k}
\end{eqnarray}
where the matrix elements $R_{\alpha \beta}$ are giving by
(\ref{R-matrix}) and
\begin{eqnarray}
p_{2,k} = \frac{1}{2} ( 1 + R_{30}~\langle
\sigma_3\rangle_k).\label{p2k}
\end{eqnarray}
It follows that the conditional entropy  given by
\begin{eqnarray}
\widetilde{S}  = \sum_{k = 0 , 1} p_{2,k} S(\rho_{2,k})
\end{eqnarray}
 can be cast in the following form
\begin{eqnarray}
\widetilde{S}  = \sum_{k = 0 , 1} p_{2,k} ~ H \bigg(\frac{1}{2} +
\frac{1}{2} \sqrt{1 - 4\det
\rho_{2,k}}\bigg).\label{condit-entrop-tilde}
\end{eqnarray}
Evidently, the quantities $p_{2,k}$ and $\det \rho_{2,k}$ can be
expressed in terms of the angular directions $\theta$ and $\varphi$
by making use of equations (\ref{val-moy-theta}), (\ref{rho2k}) and
(\ref{p2k}). Consequently, we can explicitly write $\widetilde{S} =
\widetilde{S}(\theta, \varphi )$ and perform the minimization over
the azimuthal and polar angles $\theta$ and $\varphi$. However,
there exists another elegant method to perform such an optimization.
This is presented in the following subsection.

%%%%%%%%%%%%%%%%%%%%%%%%%%%%%%%%%%%%%%%%%%%%%%%%%%%%%%%%%%%%%%%%
\subsubsection{Minimization of conditional entropy}
%%%%%%%%%%%%%%%%%%%%%%%%%%%%%%%%%%%%%%%%%%%%%%%%%%%%%%%%%%%%%%%%

To minimize the conditional entropy (\ref{condit-entrop-tilde}), we
shall use the purification method and the Koashi-Winter relation
\cite{Koachi-Winter} (see also \cite{Shi}). This relation
establishes the connection between the classical correlation of a
bipartite state $\rho_{AB} \equiv \rho_{12}$ and the entanglement of
formation of its complement $\rho_{BC} \equiv \rho_{23}$. This
connection will be clarified hereafter. The density $\rho_{12}$
(\ref{rho12-matrix}) is a two-qubit state of rank two. It decomposes
as
\begin{eqnarray}
\rho_{12} = \lambda_+ \vert \phi_+ \rangle \langle \phi_+ \vert +
\lambda_- \vert \phi_- \rangle \langle \phi_- \vert
\end{eqnarray}
where the eigenvalues $\lambda_+$ and $\lambda_-$ are given by
(\ref{lambda+-}) and the corresponding eigenstates $\vert \phi_+
\rangle$ and $\vert \phi_- \rangle$ are given by
\begin{eqnarray}
\vert \phi_+ \rangle =  \frac{1}{\sqrt{2(1+p^2)}}\bigg[ (1+p)\vert
{\bf 0} , {\bf 0} \rangle + (1-p)\vert {\bf 1} , {\bf 1} \rangle
\bigg] \qquad \vert \phi_- \rangle = \frac{1}{\sqrt{2}} \bigg[\vert
{\bf 0} , {\bf 1} \rangle +
 \vert {\bf 1} , {\bf 0} \rangle \bigg]
\end{eqnarray}
in the basis (\ref{qubit-mixed}). Attaching a qubit $3$ to the
two-qubit system $1$ and $2$, we write the purification of
$\rho_{12}$ as
\begin{eqnarray}
\vert \phi \rangle = \sqrt{\lambda_1} \vert \phi_+ \rangle \otimes
\vert {\bf 0}  \rangle +  \sqrt{\lambda_-} \vert \phi_- \rangle
\otimes \vert {\bf 1} \rangle
\end{eqnarray}
such that the whole system $123$ is described by the pure density
state $\rho_{123} = \vert \phi \rangle \langle \phi \vert $ from
which one has  the bipartite densities $\rho_{12} = {\rm Tr}_3
\rho_{123}$ and $\rho_{23} = {\rm Tr}_1 \rho_{123}$.

Suppose now that a von Neumann measurement $\{ M_0 , M_1\}$ is
performed on the subsystem $1$ (here also we need positive operator
valued measurement of rank one that is proportional the one
dimensional projector). From the viewpoint of the whole system in
the pure state $\vert \phi \rangle$, the measurement gives rise to
an ensemble for $\rho_{23}$ that we denote by
$${\cal E}_{23} = \{ p_k , \vert \phi_{23,k} \rangle \}$$
where
$$ p_k = \langle \psi \vert M_k \otimes \mathbb{I} \otimes \mathbb{I} \vert \psi \rangle\quad,
\quad \vert \phi_{23,k} \rangle \langle \phi_{23,k} \vert =
\frac{1}{p_k} {\rm Tr}_1 \bigg[ (M_k \otimes \mathbb{I} \otimes
\mathbb{I}) \vert \psi \rangle \langle \psi \vert \bigg].$$

On the other hand, from the viewpoint of the state $\rho_{12}$, the
von Neuman measurement on $1$ gives rise to the ensemble for
$\rho_2$ that it was defined previously as ${\cal E}_{2} = \{
p_{2,k} , \rho_{2,k} \}$ (note that $p_k = p_{2,k}$). It is simple
to check that the ensemble ${\cal E}_{2} $ can be induced from
${\cal E}_{23}$ by tracing out the qubit $3$, namely
$$\rho_{2,k} = {\rm Tr}_3 \bigg[\vert \phi_{23,k}  \rangle \langle \phi_{23,k}  \vert \bigg].$$
We denote by $E(\vert \phi_{23,k}  \rangle )$ the measure of
entanglement for pure states. It is given by the von Neumann entropy
of the reduced subsystem $ \rho_{2,k} = {\rm Tr}_3 (\vert
\phi_{23,k} \rangle \langle \phi_{23,k} \vert)$
$$ E (\vert \phi_{23,k}  \rangle ) = S (\rho_{2,k}).$$
It follows that the average of entanglement of formation over the
ensemble ${\cal E}_{23}$
$$\overline{E}_{23} = \sum_{k = 0, 1} p_k  E (\vert \phi_{23,k}  \rangle )$$
coincides with the conditional entropy (\ref{condit-entrop-tilde}).
At this point, it is important to note that Koachi and Winter have
pointed out that the minimum value of $\overline{E}_{23}$ is exactly
the entanglement of formation of $\rho_{23}$. Consequently, the
minimal value of the conditional entropy coincides with the
entanglement of formation of $\rho_{23}$:
\begin{equation}
\widetilde{S}_{\rm min} = E(\rho_{23}),
\end{equation}
which is easy to evaluate. Indeed, we get
\begin{equation}\label{stild-min}
 \widetilde{S}_{\rm min} = E(\rho_{23}) = H(\frac{1}{2} + \frac{1}{2} \sqrt{1 - \vert C(\rho_{23})\vert^2})
\end{equation}
where the concurrence of the density $\rho_{23}$ is
$$\vert C(\rho_{23})\vert^2 =  \frac{p^{2}(1 - p^{2})(1 - p^{2n-4})}{(1+p^{n}\cos m\pi)^2}$$
%\frac{p^{2}(1 - p^{2})(1 - q^{2})}{(1+p^{n}\cos m\pi)^2} =
 Using the
equation (\ref{condit-entrop-tilde}), one can verify the relation
$$ \widetilde{S}_{\rm min} =  \widetilde{S} (\theta = \frac{\pi}{2}, \phi = 0)$$
which reflects that the minimal value of conditional entropy is
reached for $\theta = \frac{\pi}{2}$ and $\phi = 0$. Finally,
reporting (\ref{entropy12}), (\ref{entropy1}) and (\ref{stild-min})
in the definition (\ref{def: discord}), the explicit expression of
quantum discord for the density $\rho_{12}$ is
\begin{eqnarray}\label{qdfinal-mixe}
D(\rho_{12}) = H\bigg(\frac{1}{2} \frac{(1+ p)(1 +
p^{n-1}\cos(m\pi))}{1 + p^n\cos(m\pi)}\bigg) - H\bigg(\frac{1}{2}
\frac{(1 + p^2)(1 + p^{n-2}\cos(m\pi))}{1 +
p^n\cos(m\pi)}\bigg)\label{qdmixte}\\\nonumber + H\bigg(\frac{1}{2}
+ \frac{1}{2} \sqrt{1 - \frac{p^{2}(1 - p^{2})(1 -
p^{2n-4})}{(1+p^{n}\cos m\pi)^2}})\bigg)
\end{eqnarray}
in term of the  the overlap $p$. Clearly, this result can be also
derived  using the method developed (independently) in
\cite{Mazhar10}  and \cite{Fanchini} (see also \cite{B.Li}) to
obtain the quantum discord for some special instances of the
so-called $X$ states. However, one should stress that our method
based on the purification trick combined with the Koashi-Winter
relation reduce drastically the optimization process of the
conditional entropy. It  provides a direct relation between quantum
discord and entanglement of formation reflecting how these
correlations are distributed in a given tripartite pure system. It
has been used in several papers dealing with quantum correlations as
for instance in \cite{Fanchini2} where the authors emphasized its
crucial role in exploring  the distribution of entanglement in the
deterministic quantum computation with one single pure qubit and a
collection of an arbitrary number of mixed states.

%%%%%%%%%%%%%%%%%%%%%%%%%%%%%%%%%%%%%%%%%%%%%%%%%%%%%%%%%%%%%%%%
\subsubsection{Some particular cases}
%%%%%%%%%%%%%%%%%%%%%%%%%%%%%%%%%%%%%%%%%%%%%%%%%%%%%%%%%%%%%%%%

We start with the special case $n = 2$. The state (\ref{rho12}) is
pure and $q = 1$. The quantum discord (\ref{qdmixte}) reads
$$D(\rho_{12}) = H\bigg(\frac{(1+p)(1 + p\cos m\pi)}{2(1 + p^2 \cos m\pi)}\bigg).$$
It coincides with the quantum discord $D(\rho_{k,n-k})$ given by the
equation (\ref{qdpure}) for $n = 2$ and  $k=1$. In particular, for
symmetric states ($m$ even), one obtains
$$ D(\rho_{12}) = D(\rho_{1,1})=  H\bigg(\frac{(1 + p)^2}{2(1 + p^2 )}\bigg)$$
and $ D(\rho_{12}) = D(\rho_{1,1}) = 1$ for antisymmetric states
($m$ odd) as shown in the figure 1.

Here also, it is  interesting to consider the limiting cases $p
\rightarrow 0$ and $p \rightarrow
 1$ as it was done previously for pure states. For $n > 2$, the quantum discord vanishes (\ref{qdfinal-mixe}) when $p \rightarrow
0$.  It vanishes also when $p \rightarrow 1$ for $m$ even. However,
for $m$ odd, the quantum discord (\ref{qdfinal-mixe}) reduces to the
special form
\begin{equation}
D(\rho_{12}) = H\bigg(1 - \frac{1}{n}\bigg) + H\bigg(\frac{1}{2} +
\frac{1}{2} \frac{\sqrt{n^2 - 4n + 8}}{n}\bigg) - H\bigg(1 -
\frac{2}{n}\bigg)\label{qd-Wstates}
\end{equation}
and goes to zero for $n$ large.

To corroborate  our analysis, we give in the figures 2 and 3 the
behavior of the quantum discord (\ref{qdfinal-mixe}). Figure 2 gives
a plot of quantum discord versus the overlap $p$ for the mixed state
$\rho_{12}$ with $m$ even (the symmetric case).  As seen from the
figure, after an initial increasing, the quantum discord decreases
to vanish when $p \rightarrow 1$. The maximum  of quantum discord
depends on the number of particles contained in the system. It is
remarkable that for $n = 25$ the maximum is larger than ones
obtained for $n = 4$ and $n = 5$. Beside the numerical results
reported in the figure 2, we have also studied the behavior of
quantum discord for other values of $n$. This study shows that for $
6 \leq n \leq 25 $ the maximum of quantum discord is greater than
one reached for $n = 5$. Also, the behavior of quantum discord for $
n \geq 20$ is very close to the case $n = 25$ presented in the
figure 2. In figure 3, we give a plot of the quantum discord
(\ref{qdfinal-mixe}) for $m$ odd (the antisymmetric case) and
different values of $n$. In this case the quantum discord increases
as $p$ increases for a small number of particles and the maximal
value of quantum discord is reached for $p \rightarrow 1$. However,
for a higher number of particles $n$ ($n = 25$ for instance), the
maximum is reached for $p < 1$. In the limit $p \rightarrow 1$, we
have a Werner state $\vert W \rangle_n$ and the pairwise quantum
discord given by (\ref{qd-Wstates}) decreases to vanish in the limit
of a large number of particles ($ n \rightarrow \infty $). For $p
\rightarrow 0$, corresponding to ${\rm GHZ}$ type states, the
quantum discord is zero.

\begin{center}
  \includegraphics[width=4in]{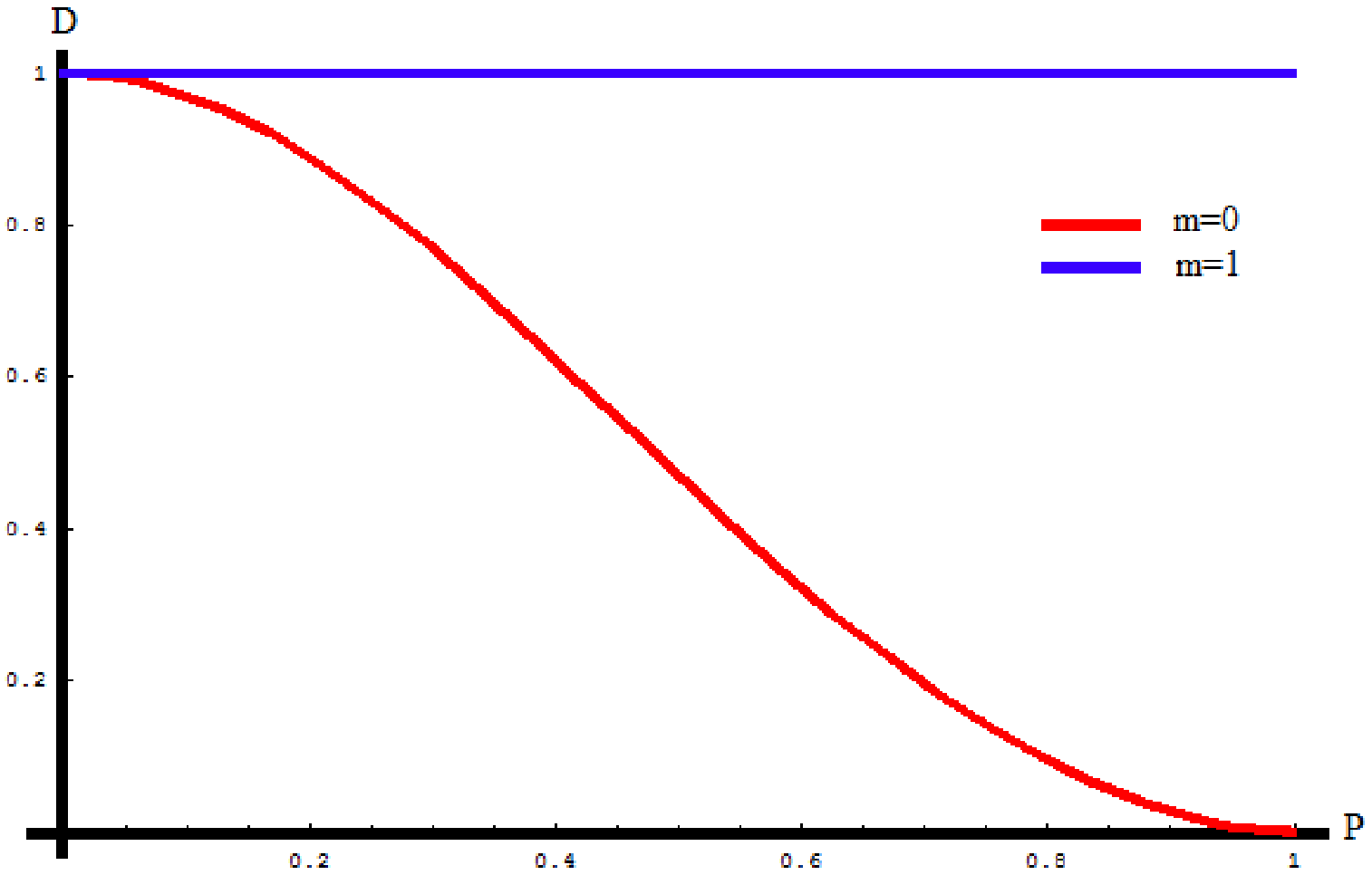}\\
FIG. 1:  {\sf The pairwise quantum discord $D$ versus the overlap
$p$ for  $n=2$.}
\end{center}
\begin{center}
  \includegraphics[width=4in]{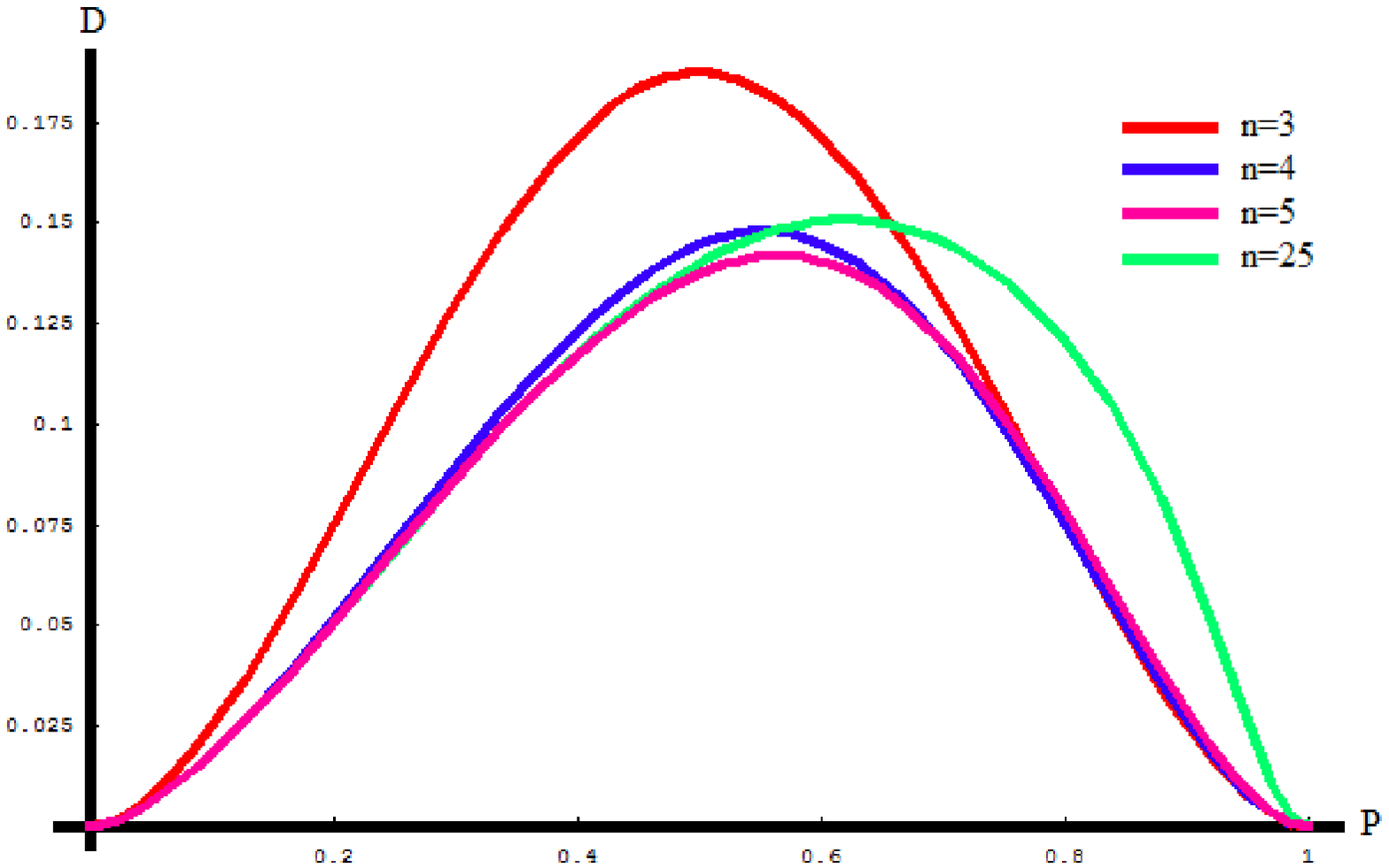}\\
FIG. 2:  {\sf The pairwise quantum discord $D$ versus the overlap
$p$ for symmetric states.}
\end{center}
\begin{center}
  \includegraphics[width=4in]{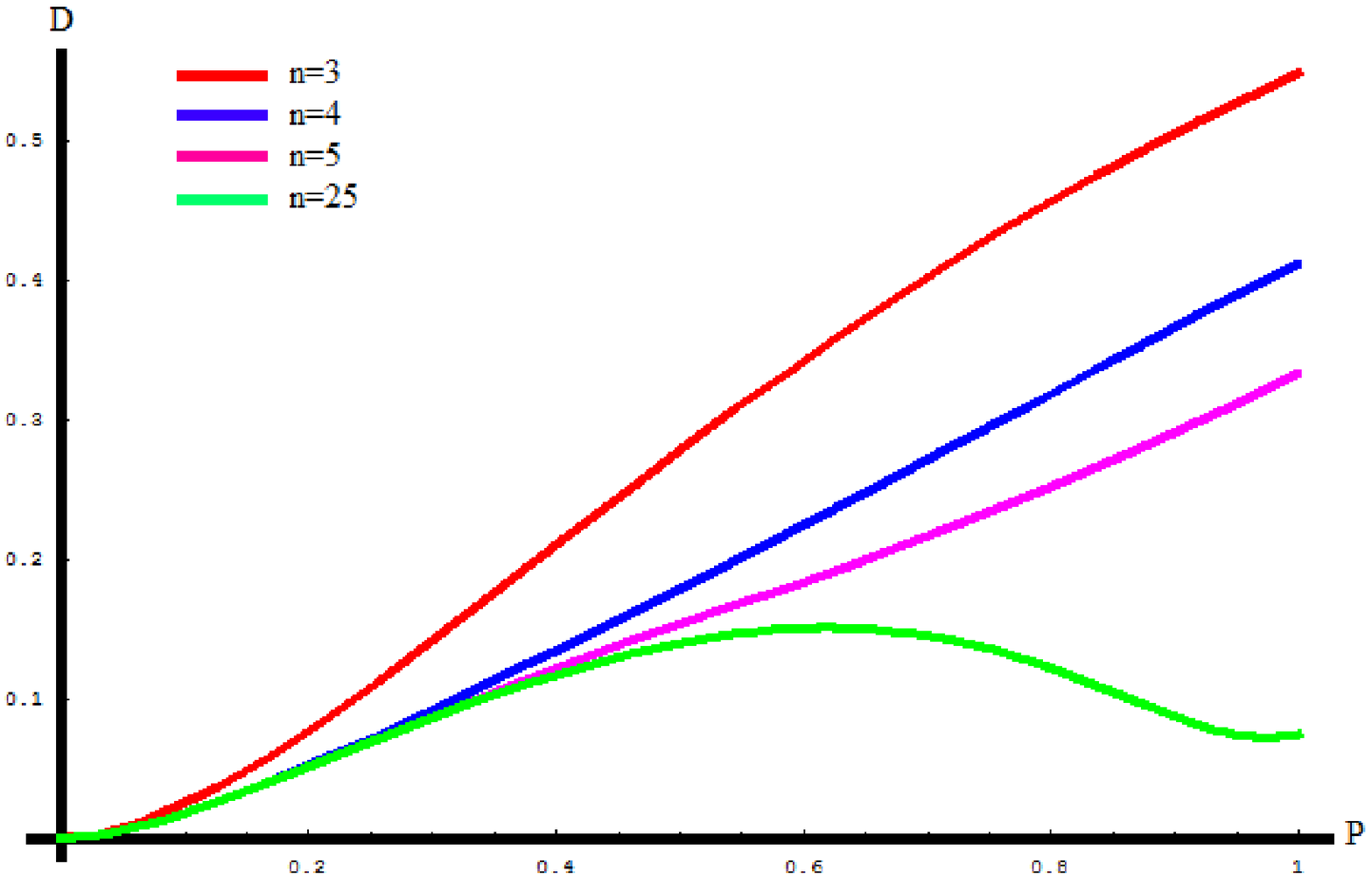}\\
FIG. 3:  {\sf The pairwise  quantum discord $D$ versus the overlap
$p$ for anti-symmetric states.}
\end{center}

%%%%%%%%%%%%%%%%%%%%%%%%%%%%%%%%%%%%%%%%%%%%%%%%%%%%%%%%%%%%%%%%%%%%%%%%%%%%%%%%%%%%%
\section{Dynamics of quantum correlations under dephasing channel}
%%%%%%%%%%%%%%%%%%%%%%%%%%%%%%%%%%%%%%%%%%%%%%%%%%%%%%%%%%%%%%%%%%%%%%%%%%%%%%%%%%%%

The sudden disappearance of entanglement is one of the most
intriguing features in quantum mechanics. In fact, it has been
observed that in a pair of entangled qubits, interacting with  noisy
environments, entanglement can disappear in a finite time \cite{Yu}.
This phenomenon, termed in the literature "entanglement sudden
death", was experimentally confirmed \cite{Almeida}.\\
In this section, we investigate the dynamics of bipartite quantum
correlations (entanglement and quantum discord) of the multipartite
coherent states $\vert z , m, n\rangle$ ($n>2$). We focus on the
second partitioning scheme (\ref{rho12}) where mixed states emerge.
The quantum discord  and entanglement are two different correlations
for mixed states (contrarily to the pure case). We consider the
evolution of a mixed bipartite system under a  dephasing dissipative
channel. In this order, we use the Kraus operator approach
\cite{NC-QIQC-2000} which describes conveniently the dynamics of two
qubits interacting independently with individual environments. The
time evolution of the bipartite density $\rho_{12} \equiv
\rho_{12}(0)$ (\ref{rho12-matrix}) can be written compactly as
$$ \rho_{12}(t) = \sum_{\mu, \nu}E_{\mu , \nu}(t) ~\rho_{12}(0)~ E_{\mu , \nu}^{\dagger}(t) $$
where the so-called Kraus operators
$$E_{\mu , \nu}(t) = E_{\mu}(t)\otimes E_{\nu}(t) \qquad \sum_{\mu,
\nu}E_{\mu , \nu}^{\dagger}  E_{\mu , \nu} = \mathbb{I}.$$ The
operators $E_{\mu}$ describe the one-qubit quantum channel effects.
The non-zero Kraus operators for a dephasing channel are given by
$$ E_0 = {\rm diag}( 1 , \sqrt{1 - \gamma})  \qquad  E_1 = {\rm diag}( 0 , \sqrt{\gamma})$$
with $ \gamma = 1 - e^{-\Gamma t}$ and $\Gamma$ denoting the decay
rate. It is easy  to check that the density matrix
(\ref{rho12-matrix}) evolves as
\begin{equation}
\rho_{12} (t) = {\cal N}^2 \left( \begin{smallmatrix} 2a^4(1+q\cos
m\pi)& 0
    & 0& 2 (1 -
\gamma)a^2b^2(1+q\cos m\pi
)\\
0  & 2a^2b^2(1-q\cos m\pi ) & 2 (1 - \gamma)a^2b^2(1-q\cos m\pi
) & 0 \\
0  & 2 (1 - \gamma)a^2b^2(1-q\cos m\pi ) & 2a^2b^2(1-q\cos m\pi
) & 0 \\
2 (1 - \gamma)a^2b^2(1+q\cos m\pi ) & 0 & 0 & 2b^4(1+q\cos m\pi)
\end{smallmatrix}
\right)
\end{equation}
Using the prescription provided in the works \cite{Wootters98} and
\cite{Hil97} to measure the amount of entanglement in bipartite
quantum states, one can check that the concurrence is  given by
$$ C(t) = 2~ {\rm max} \{ 0 , \Lambda_1(t) , \Lambda_2(t) \}$$
where
$$ \Lambda_1(t) = 2 {\cal N}^2  a^2b^2 \bigg[   (1 - \gamma)(1+q\cos m\pi ) - (1-q\cos m\pi )\bigg],$$
$$\Lambda_2(t) = 2 {\cal N}^2  a^2b^2 \bigg[   (1 - \gamma)(1-q\cos m\pi ) - (1+q\cos m\pi )\bigg].$$
It follows that  the concurrence is given by
$$ C (t)= \frac{1}{2}~ \frac{1 - p^2}{1 + p^n\cos m\pi}\bigg[ e^{-\Gamma t}(1 + p^{n-2}) - (1-p^{n-2})\bigg]$$
for
$$ t < t_0 = \frac{1}{\Gamma} [ ~\ln(1 + p^{n-2}) -  \ln(1 - p^{n-2})]$$
and the system is entangled. However, for $ t > t_0$, the
concurrence is zero and the entanglement disappears, i.e. the system
is separable. This clearly reflects  that under  dephasing channel,
the entanglement suddenly vanishes. Note that the bipartite system
under consideration is initially (in the absence of an external
interaction) entangled. Indeed, for $ t = 0$, the concurrence  is
$$C(0) =  \frac{p^{n-2} - p^n}{1 + p^n\cos m\pi},$$
and  is always non zero except in the limiting cases $ p \rightarrow
0$ or $ p \rightarrow 1$ for $m$ even. Notice that for $m$ odd, the
concurrence is zero for $p \rightarrow 0$  but does not vanish when
$ p \rightarrow 1$ and it is given by $2/n$.

It is important to stress that the quantum discord $D(\rho_{12}(0))$
is nonzero except in the particular case $ p \rightarrow 0$. To show
that the quantum discord does not disappear after the interaction of
the system with the dissipative channel, we note that the density
matrix $\rho_{12}(t))$ belongs to the class of the so-called
circulant states \cite{Bylicka}. Thus, one can use the vanishing
quantum discord criteria, discussed in  \cite{Bylicka}, to check
that the state $\rho_{12}(t)$ has vanishing quantum discord if and
only if $ p \rightarrow 0$. This implies that even when entanglement
suddenly disappears in a finite time, quantum discord does not
vanish. This agrees with the commonly accepted fact that the quantum
discord is more robust than entanglement to sudden death under a
dissipative channel.

%%%%%%%%%%%%%%%%%%%%%%%%%%%%%%%%%%%%%%%%%%%%%%%%%%%%%%%%%%%%%%%%%%%%%%%%%%%%%%%%%%%%%%%%%%%
\section{ Concluding remarks}
%%%%%%%%%%%%%%%%%%%%%%%%%%%%%%%%%%%%%%%%%%%%%%%%%%%%%%%%%%%%%%%%%%%%%%%%%%%%%%%%%%%%%%%%%%%

In this paper, we have obtained explicit expressions of quantum
discord for symmetric and antisymmetric superpositions of
multipartite coherent states. These states cover the generalized
${\rm GHZ}$ and generalized Werner states. In particular, the
balanced antisymmetric superpositions ($m$ odd) interpolate
continuously between generalized ${\rm GHZ}$ and generalized Werner
states. The key point in determining the bipartite quantum
correlations is based on the splitting of the entire system in two
qubit subsystems. Two inequivalent splitting schemes were discussed.
The first one leads to a pure bipartite density and the quantum
discord coincides with the entanglement of formation. The second
consists in constructing bipartite systems by a trace procedure
keeping only the modes in which we are interested. In this way,
mixed states are obtained and the corresponding quantum discord was
explicitly derived. This derivation  requires an optimization over
all the measurement needed to extract the amount of quantum
correlation which is general difficult to perform. To avoid such a
difficulty, we used the purification method together with  the
Koashi-Winter relation which are advantageous in simplifying the
minimization process of the conditional entropy. In the last part of
the paper, we discussed the robustness of the quantum discord
present in multipartite coherent states in comparison with the
entanglement. We have shown that in sending the system through a
dephasing channel the entanglement can be lost. This is not the case
of quantum discord which  behaves more robust against dissipative
channels. It must be noticed that the results obtained here can be
extended easily to many other classes of coherent states even those
associated with higher symmetries which are labeled by several
variables. Finally, we stress that the bipartite correlation does
not capture genuine multipartite correlations. It follows that it is
interesting to investigate the measure of genuine multipartite
quantum discord for arbitrary multipartite non orthogonal states in
the spirit of the results recently obtained in \cite{Ma}.


\begin{thebibliography}{99}

\bibitem{Horodecki-RMP-2009} R. Horodecki, P. Horodecki, M. Horodecki and K. Horodecki, Rev. Mod. Phys. {\bf 81}(2009) 865.

\bibitem{Guhne} O. G\"uhne and G. T\'oth, Phys. Rep. {\bf 474} (2009) 1.

\bibitem{vedral} K. Modi, A. Brodutch, H. Cable, T. Paterek and V. Vedral, {\rm Quantum discord and other measures of quantum
correlation}, {\tt arXiv:1112.6238}.

\bibitem{NC-QIQC-2000} M.A. Nielsen and I.L. Chuang, {\it Quantum Computation and Quantum Information} (Cambridge Univ. Press, Cambridge,
2000).

\bibitem{Vedral-RMP-2002} V. Vedral, Rev. Mod. Phys. {\bf 74} (2002)
197.

\bibitem{Rungta}  P. Rungta, V. Buzek, C.M. Caves, M. Hillery and G.J. Milburn, Phys. Rev. A {\bf 64} (2001) 042315.

\bibitem{Ben3}  C.H. Bennett, D.P. DiVincenzo, J. Smolin and W.K.
Wootters, Phys. Rev. {\bf A 54} (1996) 3824.

\bibitem{Wootters}  W.K. Wootters, Phys. Rev. Lett. {\bf 80} (1998) 2245.

\bibitem{Coffman}  V. Coffman, J. Kundu and W.K. Wootters, Phys. Rev.
A {\bf 61} (2000) 052306.

\bibitem{Knill} E. Knill and R. Laflamme, Phys. Rev. Lett. {\bf 81} (1998)
5672.

\bibitem{Braunstein-PRL83-1999} S.L. Braunstein, C.M. Caves, R. Jozsa, N. Linden, S. Popescu and R. Schack, Phys. Rev. Lett. {\bf 83} (1999)
1054.

\bibitem{Bennett-PRA59-1999} C.H. Bennett, D.P. DiVincenzo, C.A. Fuchs, T. Mor, E. Rains, P.W. Shor, J.A. Smolin and W.K. Wootters,
Phys. Rev. A {\bf 59} (1999) 1070.

\bibitem{Meyer-PRL85-2000} D.A. Meyer, Phys. Rev. Lett. {\bf 85} (2000) 2014.

\bibitem{Biham}  E. Biham, G. Brassard, D. Kenigsberg and T. Mor, Theor. Comput. Sci.
320 (2004) 15.

\bibitem{Datta-PRL100-2008} A. Datta, S.T. Flammia and C.M. Caves, Phys. Rev. A {\bf 72}, 042316 (2005);
A. Datta and G. Vidal, {\it ibid}. {\bf 75}(2007) 042310; A. Datta,
A. Shaji and C.M. Caves, Phys. Rev. Lett. {\bf 100} (2008) 050502.

\bibitem{Vedral-et-al} L. Henderson and V. Vedral, J. Phys. A {\bf 34}(2001)  6899;
V. Vedral, Phys. Rev. Lett. {\bf 90}  (2003) 050401; J. Maziero, L.
C. Cel\'eri, R.M. Serra and V. Vedral, Phys. Rev A {\bf 80} (2009)
044102.

\bibitem{Ollivier-PRL88-2001} H. Ollivier and W.H. Zurek, Phys. Rev. Lett. {\bf 88} (2001) 017901.

\bibitem {Luo08} S. Luo, Phys. Rev. A \textbf{77} (2008) 042303; Phys.
Rev. A \textbf{77} (2008) 022301.

\bibitem {Mazhar10} M. Ali, A.R.P. Rau and G. Alber, Phys. Rev. A \textbf{81} (2010)
042105.

\bibitem{Giorda} P. Giorda and M. G. A. Paris, Phys. Rev. Lett. {\bf 105} (2010)
020503.

\bibitem{Dakic} B. Dakic, Y.O. Lipp, X. Ma, M. Ringbauer, S. Kropatschek, S. Barz,
T. Paterek, V. Vedral, A. Zeilinger, C. Brukner and P. Walther,
Quantum discord as optimal resource for quantum communication, {\tt
arXiv:1203.1629} (2012).

\bibitem{Sanders3} B.C Sanders, J. Phys. A: Math. Theor. {\bf 45} (2012) 244002.

\bibitem{Sanders} B.C. Sanders, Phys. Rev. A 45 (1992) 6811.

\bibitem{Wang} X. Wang, B. C. Sanders and S. H. Pan, J. Phys. A 33 (2000)
7451.

\bibitem{Sanders2}  B.C. Sanders, Phys. Rev. A {\bf 46} (1992) 2966.

\bibitem{Jex} I. Jex, P. T\" orm\" a and S. Stenholm, J. Mod. Opt. {\bf 42} (1995)
1377.

\bibitem{Zheng} S.-B. Zheng, Quant. Semiclass. Opt. B: J. European Opt. Soc. B {\bf 10} (1998) 691 .

\bibitem{Wang1} X. Wang and B.C. Sanders, Phys. Rev. A {\bf 65} (2001) 012303.

\bibitem{Jeong1} H. Jeong and N. B. An, Phys. Rev. A {\bf 74} (2006) 022104 .

\bibitem{Li}
H.-M. Li, H.-C. Yuan and H.-Y. Fan, Int. J. Theor. Phys. {\bf 48}
(2009) 2849.

\bibitem{Munhoz} P. P. Munhoz, F. L. Semi\~ao and Vidiello, Phys. Lett. A {\bf
372} (2008) 3580.

\bibitem{Wang-WF} W.-F. Wang, X.-Y. Sun and X.-B. Luo, Chin. Phys. Lett. {\bf
25}(2008) 839.

\bibitem{Becerra} E.M. Becerra-Castro, W.B. Cardoso, A.T. Avelar and B. Baseia, J. Phys. B: At. Mol. Opt. Phys. {\bf 41} (2008) 085505.

\bibitem{Milburn1} G.J. Milburn, Phys. Rev. A {\bf 33} (1985) 674.

\bibitem{Milburn2}  G.J. Milburn and C.A. Holmes, Phys. Rev. Lett. {\bf 56} (1986) 2237.

\bibitem{Yurke1}  B. Yurke and D. Stoler, Phys. Rev. Lett. {\bf 57} (1986) 13.

\bibitem{Yurke2}  B. Yurke and D. Stoler, Phys. Rev. A {\bf 35} (1987) 4846.

\bibitem{Buzek} V. Bu$\tilde{z}$ek and P. L. Knight, Progress in Optics {\bf 34} (1995) 1.

\bibitem{Brune} M. Brune, E. Hagley, J. Dreyer, X. Maitre, A. Maali, C. Wunderlich, J. M. Raimond and S. Haroche, Phys. Rev. Lett. {\bf 77} (1996) 4887.

\bibitem{Perelomov} A. Perelomov, {\it Generalized Coherent States and Their Applications},
Springer, Berlin, 1986.

\bibitem{daoud} M. Daoud and M. Kibler, {\it Bosonic and k-fermionic coherent states for a class of polynomial Weyl-Heisenberg
algebras}, {\tt arXiv:1110.4799}, To appear in J. Phys. A (2012).

\bibitem{Dur00} W. D\"ur, G. Vidal and J.I. Cirac, Phys. Rev. A
{\bf 62} (2000) 062314.

\bibitem{Ourjoumtsev} A. Ourjoumtsev, H. Jeong, R. Tualle-Brouri and P. Grangier, Nature
{\bf 448} (2007) 784.

\bibitem{Gerrits} T. Gerrits, S. Glancy, T.S. Clement, B. Calkins, A.E. Lita, A.J.
Miller, A. L. Migdall, S.W. Nam, R.P. Mirin and E. Knill, Phys. Rev.
A {\bf 82} (2010) 031802(R).

\bibitem{Hamieh} S. Hamieh, R. Kobes and H. Zaraket, Phys. Rev. A {\bf 70}
(2004) 052325.

\bibitem{Galve} F. Galve, G. Giorgi and  R. Zambrini, EPL {\bf 96} (2011) 40005

\bibitem{Modi} K. Modi, A. Brodutch, H. Cable, T. Paterek and V.
Vedral,  Quantum discord and other measures of quantum correlation,
{\tt arXiv:1112.6238}.

\bibitem{Koachi-Winter} M. Koachi and A. Winter, Phys. Rev. A {\bf
69} (2004) 022309.

\bibitem{Shi} M. Shi, W. Yang, F. Jiang and J. Du,  J. Phys. A: Math. Theor. {\bf 44} (2011)
415304.

\bibitem{Fanchini} F.F. Fanchini, T. Werlang, C.A. Brasil, L.G.E. Arruda and A.O.
Caldeira, Phys. Rev. A. {\bf 81} (2010) 052107.

\bibitem{B.Li} B. Li, Z-X Wang and S-M Fei, Phys. Rev. A {\bf 83} (2011) 022321.

\bibitem{Fanchini2} F.F. Fanchini, M.F. Cornelio, M.C. de Oliveira and A.O. Caldeira,
Phys. Rev. A {\bf 84} (2011) 012313.

\bibitem{Yu} T. Yu and J.H. Eberly, Phys. Rev. Lett. {\bf 97} (2006) 140403 .

\bibitem{Almeida} M.P. Almeida, F. de Melo, M. Hor-Meyll, A. Salles, S.P. Walborn,
P.H. Souto Ribeiro and L. Davidovich, Science {\bf 316} (2007) 579.


\bibitem{Wootters98} W.K. Wootters, Phys. Rev. Lett. {\bf 80} (1998) 2245; W.K. Wootters,  Quant. Inf. Comp. {\bf 1} (2001)
27.

\bibitem{Hil97} S. Hill and W.K. Wootters, Phys. Rev. Lett. {\bf 78} (1997)
5022.


\bibitem{Bylicka} B. Bylicka, D. Chru\'sci\'nski, Circulant states with vanishing
quantum discord, {\tt arXiv:1104.1804}.

\bibitem{Ma} Z-H. Ma and Z-H. Chen, Witness for a measure of genuine
multipartite quantum discord for arbitrary $N$ partite quantum
state, {\tt ArXiv:1108.4323}

%%%%%%%%%%%%%%%%%%%%%%%%%%%%%%%%%%%%%%%%%%%%%%%%%%%%%%%%%%%%%%%%%%%%%%%%%%%%%%%%%%%%%%%%%%%%%%%





\end{thebibliography}
\end{document}